\documentclass[12pt]{article}

\input epsf
\begin {document}
\begin{titlepage}
\begin{flushright}
{\small
NYU-TH-05/02/6}
\end{flushright}
\vskip 0.9cm

\centerline{\Large \bf 
Strings at the bottom of the deformed conifold}
\vspace{0.5cm}
\centerline{\Large \bf 
 }

\vskip 0.7cm
\centerline{\large Jose J. Blanco-Pillado\footnote{E-mail: blanco-pillado@physics.nyu.edu} 
and Alberto Iglesias\footnote{E-mail: iglesias@physics.nyu.edu}}
\vskip 0.3cm
\centerline{\em Center for Cosmology and Particle Physics}
\centerline{\em Department of Physics, New York University, New York, 
NY, 10003, USA}                                                          

\vskip 1.9cm 

\begin{abstract}
We present solutions of the equations of motion of macroscopic F and D strings
extending along the non compact 4D sections of the conifold geometry and 
winding around the internal directions. The effect of the Goldstone modes 
associated with the position of the strings on the internal manifold can be 
seen as a current on the string that prevents it from collapsing and allows the
possibility of static 4D loops. Its relevance in recent models of brane 
inflation is discussed.

\end{abstract}

\end{titlepage}

\newpage

\section{Introduction}

{}There has been a considerable amount of effort recently in trying to embed
inflation within the context of string theory. All these models are based on 
the simple realization that moduli associated with the ingredients necessary 
to go from ten to four dimensions will be seen as light fields in the four 
dimensional description. In this regard, we can distinguish between two 
different origins for the would be inflaton.

{}On the one hand, compactifications mechanisms generate many moduli
that come from different components of the metric, the dilaton, or the 
different antisymmetric forms present in the low energy description of string 
theory in ten dimensions. But these moduli should be fixed today by their 
respective potentials, so one of the most simple frameworks to get inflation 
would be to use these potentials to create a period of inflationary expansion.
This idea has been pursued within string theory for a long time 
\cite{modular} and has its most recent incarnation in \cite{racetrack}.

{}On the other hand, the discovery of D-branes has opened up a new opportunity 
to get inflation within string theory \cite{gia1,gia2,fernando}. The idea is 
that we can use the moduli that parametrizes the distance 
between the branes along the compact directions as the inflaton field. In 
\cite{KKLMMT} these type
of models were discussed within the framework of compactification with fluxes 
\cite{GKP}. The authors argue that
due to the potentials for the D-branes coming from the background geometry, it 
would be natural to have this process of inflation happening when the branes 
are located on a warped region of the internal 
manifold. These regions, called throats can be thought as regularized 
deformations of the conifold singularity \cite{CO} on 
a Calabi-Yau. In \cite{KS} the authors were 
able to find an exact solution of the supergravity equations of motion that 
describe the geometry once some fluxes
are turned on along the cycles of the internal space. We will review 
this solution in section III of this paper since it is important for our purposes.

{}Brane inflation ends in these scenarios when the branes collide
at the bottom of this warped geometry. It has been suggested that brane 
annihilation would leave behind a network of lower dimensional extended 
objects \cite{Tye, AlexGia, CMP, Polchinski}, which would be seen 
as strings from the
four dimensional point of view. This alternative has renewed the interest 
on the possibility of observing cosmological consequences of cosmic strings,
either fundamental strings (F strings) or D1-branes (D strings).  

On the other hand, many high energy extensions of the standard model predict 
that our universe underwent a phase transition during which one dimensional topological 
defects of the field theory in question could have been produced \cite{Kibble, Alex-report}.
These would be the usual cosmic strings that have been extensively studied 
in the past in relation with different cosmological and astrophysical 
observations \cite{Alex-book, Kibble-Hindmarsh}.
It is therefore interesting to look for distinguishing features between these 
two types of cosmic string networks. Some suggestions in this regard have been put 
forward in the literature \cite{Jackson, Damour-Alex, Shellard}. 

Here we explore the possibility that these strings were able to propagate 
in the extra dimensional part of the geometry. It is clear that this could have 
important consequences for the evolution of the network of strings \cite{Shellard}.
In this paper we demonstrate this fact by finding stable solutions for 4-dimensional
extended closed loops that wind around a circle of the internal space. 
We show that even though the extra dimensional part of the internal geometry in this 
region is a three sphere threaded with fluxes, the solutions remain stable for some 
range of the parameters. On the other hand, these 
light degrees of freedom that parametrize the position of the string on the compact space
can also be thought of as a neutral current
flowing along the string. This current has a very peculiar equation of state that
could in principle help us to distinguish between fundamental versus field theory
cosmic strings.

{}The organization of the paper is the following. In section II we present the 
simple solutions of strings propagating in $M^4  \times S^1$, and discuss its 
connection to the superconducting string case. Section III discusses the 
embedding and stability of these solutions in the deformed conifold geometry. 
Finally in the conclusions we elaborate on the possible role that these 
solutions may play within the recently proposed scenarios of brane inflation.

\section{Strings on $M^4  \times S^1$}

{}The Nambu-Goto action of a string propagating in D+1 dimensions is given by,
\begin{equation}
S = -{\cal T} \int{\sqrt{-\gamma}~d\sigma dt}~,
\end{equation}
where ${\cal T}=1/2\pi\alpha^\prime$ is the tension of the string, 
$\gamma_{ab}= g_{MN} \partial_a X^M 
\partial_b X^N$ is the induced metric on the worldsheet parametrized by the 
intrinsic coordinates, $\sigma$ and $t$, and $X^M(\sigma,t)$
gives the embedding of the string motion in the D+1-dimensional space-time. The
equations of motion in the conformal gauge for a string propagating in a 
spacetime with metric $g_{MN}$ are \cite{Alex-book},
\begin{equation}\label{eom}
\ddot{X}^M- {X''}^M+ \Gamma^{M}_{NP}(\dot{X}^N \dot{X}^P - {X'}^N {X'}^P)=0
\end{equation}
where $X'$ and $\dot{X} $ denote the derivatives with respect to $\sigma$ and 
$t$ and $\Gamma^{M}_{NP}$ are the Christoffel symbols for the background metric. On 
the other hand, 
in order to be in the conformal gauge, we have to impose on the solutions the 
following constraints:
\begin{eqnarray}
\dot{X}^M  {X'}_M = 0  ~,\\
\dot{X}^M \dot{X}_M + X'^M X'_M = 0~.
\end{eqnarray}

We can now particularize these equations to the case at hand, a 4+1 dimensional
spacetime with one of the spatial dimensions compactified on a circle. In this 
case, the metric is:
\begin{equation}
ds^2= g_{MN} dX^M dX^N = -dt^2 + dx_i dx^i + d\theta^2.
\end{equation}

{}In this background, we can use the extra gauge freedom to choose $X^0=t $ 
so that the 4-dimensional spatial vector $X^J$ 
that parametrizes the position of the string satisfies the following set of 
equations:
\begin{eqnarray}
\ddot{X}^J={X^J}'' ~,\\
\dot{X}^J  {X'}_J = 0 ~, \\
\dot{X}^J \dot{X}_J + X'^J X'_J = 1~.
\end{eqnarray}

We look for solutions of these equations that describe macroscopic strings extended
along the non compact 3+1 dimensional part of
this spacetime and winding the transverse circle, namely solutions of the following form,
\begin{eqnarray}
x^0(\sigma,t) &=& t ~,\\
{\bf x}(\sigma,t) &=& {\bf a}(\sigma-t)~, \\
\theta(\sigma,t) &=& \sigma + (\sigma - t) b~,
\end{eqnarray}
where ${\bf a}$ is an arbitrary vector function and $b$ a constant.
It is clear that this ansatz fulfills the equations of motion, so we only 
have to impose the constraint equations which are also fulfilled provided 
that,
\begin{equation}
b(b+1)+|{\bf a}'|^2=0~.
\end{equation}
Note that we also have to choose $b$ such that it respects the periodicity of
$\theta$. 

{}This solution describes a wiggle of arbitrary shape propagating on a 
straight string along the extra dimension. Solutions of this type have been known in 
the cosmic string literature for quite some time \cite{waves} and generalize the 
solutions found in the circular case in \cite{Wali}. They have also been discussed 
recently within string theory in \cite{tubes}.  From the 
4-dimensional perspective, this string looks like a loop of arbitrary form 
stabilized by the backreaction on its worldsheet of the perturbations of the string
along the extra dimension. In this regard, they are basically a compactified 
version of the wiggly cosmic
string model \cite{Carter-wiggly, Alex-wiggly}.

We can also think of these wiggles along the compactified directions as an induced current 
on the worldsheet \cite{Nielsen}, and therefore consider these strings as a model of neutral
superconducting strings \cite{Witten}. Hence, it is not so surprising that 
there are stable configurations of string loops since states like this are 
well known in the context of superconducting cosmic string models
\cite{Davis}. Also the arbitrary shape for the strings in this case is easily understood
due to the fact that the 4-dimensional string tension vanishes in 
those configurations \cite{chiral1,chiral2}.

\section{Strings on the deformed conifold}

{}We now consider solutions of the type described in the previous section propagating
 in the background of the warped deformed conifold of 
type $IIB$ supergravity \cite{KS}. In this solution the dilaton field is 
constant, ${\rm e}^\Phi=g_s$, and the line element, 3-form   
and self-dual 5-form Ramond-Ramond field strengths, and the Kalb-Ramond 2-form 
are given by
\begin{eqnarray}
ds_{10}^2 &=& h^{-1/2}(\tau)dx^\mu dx_\mu + {1\over 2} h^{1/2}(\tau)
\epsilon^{4/3}
K(\tau)\left[{1\over 3K^3(\tau)}\left(d\tau^2+g_5^2\right)+\right.\\
&& \left. {\rm cosh}^2\left({\tau
\over 2}\right)\left(g_3^2+g^2_4\right)+{\rm sinh}^2\left({\tau\over 2}\right)
\left(g_1^2+g_2^2\right)\right]~,\\
F_3&=&{1\over 2}M\alpha'\{g_5\wedge g_3
\wedge g_4+d\left[{{\rm sinh}~\tau -\tau \over 2 ~{\rm sinh}~\tau}
\left(g_1\wedge g_3+g_2
\wedge g_4\right)\right]\}~,\\
\tilde F_5&=&{\cal F}_5+\star {\cal F}_5~,~~~~~~~~{\cal F}_5=B_2\wedge F_3~,\\
B_2&=&g_s M \alpha' {\tau~{\rm coth}~\tau-1\over 4~ {\rm sinh}~\tau}
d\tau\wedge\left(({\rm cosh}~\tau-1)g_1\wedge g_2+
({\rm cosh}~\tau+1)g_3\wedge g_4\right)~,
\end{eqnarray} 
where
\begin{eqnarray}
K(\tau)&=&{({\rm sinh}(2\tau)-2\tau )^{1/3}\over 2^{1/3} {\rm sinh}~\tau}~,\\
h(\tau)&=&(g_s M \alpha')^22^{2/3}\epsilon^{-8/3}\int_\tau^\infty dx~{x~{\rm 
coth} ~x -1\over {\rm sinh}^2x}({\rm sinh}(2x)-2x)^{1/3}~,
\end{eqnarray}
and the 1-forms ${g_1,\dots g_5}$ are linear combinations of the Cartan 1-forms
on the coset $SU(2)\times SU(2)/U(1)$ \cite{MT}, namely:
\begin{eqnarray}
g_1={1\over \sqrt{2}}(e_1-e_3)~,&& g_2={1\over \sqrt{2}}(e_2-e_4)~,\\
g_3={1\over \sqrt{2}}(e_1+e_3)~,&& g_4={1\over\sqrt{2}}(e_2+e_4)~,~~~~g_5=e_5~,
\end{eqnarray}
\begin{eqnarray}
&&e_1=-{\rm sin} \theta_1 d\phi_1~,~~~~ e_2=d\theta_1~,\\
&&e_3={\rm cos}\psi {\rm sin}\theta_2 d\phi_2-{\rm sin}\psi d\theta_2~,\\
&&e_4={\rm sin}\psi {\rm sin}\theta_2 d\phi_2+{\rm cos}\psi d\theta_2~,\\
&&e_5=d\psi + {\rm cos}\theta_1d\phi_1+{\rm cos}\theta_2 d\phi_2~. 
\end{eqnarray}
In this parametrization, $0\le\theta_i<\pi$, $0\le\phi_i<2\pi$, $0\le\psi<4\pi$ 
for $i=1,2$.
 
{}We are interested in the dynamics of strings near the bottom of the conifold,
thus we consider the $\tau\to 0$ limit of the supergravity solution. The
metric becomes (choosing the deformation parameter $\epsilon=12^{1/4}$)
\begin{eqnarray}\label{met2}
ds_{10}^2 &=& k^{-1}\left(1+{\tau^2\over a_0 6^{4/3}}\right)dx^\mu dx_\mu + 
k\left[{1\over 2}\left(1+\left({1\over 5}-
{1\over a_0 6^{4/3}}\right)\tau^2\right)\left(d\tau^2+g_5^2\right)+\right.
\nonumber\\
&& \left. \left(1+\left({3\over 20}-{1\over a_0 6^{4/3}}\right)\tau^2\right)
\left(g_3^2+g^2_4\right)+{\tau^2\over 4}
\left(g_1^2+g_2^2\right)\right]+{\cal O}(\tau^3)~,
\end{eqnarray}
where $k=a_0^{1/2}6^{-1/3}g_s M \alpha'$ and $a_0\approx 0.718$.

{}At $\tau=0$ the angular part of the conifold degenerates to a round three-
sphere of radius $k^{1/2}$,
\begin{equation}\label{s3}
d\Omega^2_3={1\over 2}g_5^2+g_3^2+g_4^2~,
\end{equation}
plus a collapsed two-sphere,
\begin{equation}\label{s2}
d\Omega_2^2=g_1^2+g_2^2~.
\end{equation}
The stability group of the $SU(2)\times SU(2)$ symmetric solution
of the conifold defining equations is 
enhanced from $U(1)$ to a full $SU(2)$ at $\tau=0$ \cite{CO} which we use 
to set $\theta_2=\phi_2=0$. Therefore, it is easy to see that (\ref{s3}) is 
indeed the metric on a round three-sphere as 
$d\Omega_3^2=Tr~dT^\dagger dT/2$ for the $SU(2)$ matrix 
\begin{equation}
T=\left(\begin{array}{cc} 
{\rm cos}{\theta_1\over 2}{\rm e}^{i{\psi+\phi_1\over 2}} & 
{\rm sin} {\theta_1\over 2}{\rm e}^{i{-\psi+\phi_1\over 2}} \\
-{\rm sin} {\theta_1\over 2}{\rm e}^{i{\psi-\phi_1\over 2}} &
{\rm cos}{\theta_1\over 2}{\rm e}^{-i{\psi+\phi_1\over 2}} \end{array}\right)
\end{equation} 
and (\ref{s2}) is just $d\theta_1^2+{\rm sin}^2\theta_1 d\phi_1^2$.

We want to study strings propagating in this background which form closed loops
in $I\!\!R^{3,1}$ and wind around a maximal circle on the blown-up 
three-sphere.
We parametrize the string by target space coordinates $(x^0,
\dots, x^3,\tau,\theta_1,\psi,\phi_1,\theta_2,\phi_2)$ and we use the 
worldsheet gauge freedom to fix $X^0=t$ and $(X^1/X^2)={\rm tan}~\sigma$. 
Within this gauge we take the following ansatz: 
\begin{equation}\label{ans}
X^M=(t, r(t)~{\rm sin}\sigma, r(t)~{\rm cos}\sigma ,z_0 , \tau, 0, 
2 n_\psi \sigma+\varphi(t), n_\phi \sigma+\varphi(t), 0, 0 )~,
\end{equation}
where $n_\psi$ and $n_\phi$ are integers labeling the winding in the $\psi$ and 
$\phi$ directions respectively (for $0\le\sigma<2\pi$, $2n_\psi\sigma$ runs 
$n_\psi$ times over the range of $\psi$).
Let us mention that circular string solutions winding around transverse spheres are
known \cite{DW,DLN} albeit not for the case of the warped background 
considered here.  

{}In what follows we will focus on solutions with fixed radius $r$ and find 
the conditions for their stability for the case of both a fundamental string 
and a D-string.

\subsection{F-string}

{}We consider the Nambu-Goto action for the bosonic sector of the F-string:
\begin{equation}
S=-{\cal T}\int d\sigma dt ~\sqrt{-\gamma}+\mu\int B_2~,
\end{equation}
where the charge $\mu={\cal T}$ and $B_2$ stands for the Kalb-Ramond two 
form. Two conserved quantities follow from the energy momentum 
tensor namely,\footnote{Note that $B_2$ vanishes at the bottom of the conifold.}
\begin{eqnarray}
E&=&2\pi\alpha'k\left({\delta{\cal L}\over\delta \dot X^I}\dot X^I-
{\cal L}\right)={r^2+{1\over 2}k^2s^{\prime 2}\over 
\sqrt{(1-\dot r^2)\left(r^2+{1\over2}k^2
s^{\prime 2}\right)-{1\over 2}k^2 r^2\dot s^2}}~,\label{E}\\
\ell&=&2\pi\alpha' k{\delta{\cal L}\over\delta\dot X^I}(X^I)'=
{{1\over 2}k^2 s^\prime\dot s r^2\over \sqrt{(1-\dot r^2)\left(r^2+{1\over 2}
k^2 s^{\prime 2}\right)
-{1\over 2}k^2 r^2\dot s^2}}~,\label{ell}
\end{eqnarray}
where the index $I$ labels the eight coordinates not fixed by the gauge 
choice and we have defined for convenience\footnote{The 
fact that $s/2$ plays the role of the angular coordinate on the $S^1$ case, and 
therefore that the solution should be periodic in this variable, implies that 
only even $n_\phi$ should be considered.} $s\equiv\psi+\phi_1$. 
The r.h.s. of (\ref{E}) and (\ref{ell}) are obtained by specializing to 
the ansatz (\ref{ans}).
These two conserved quantities imply that the circular orbits have radius 
$r=\sqrt{\ell}$ and transverse angular velocity 
$\dot\varphi=1/\sqrt{2}k$ for any non vanishing winding numbers $n_\psi,
~n_\phi$. It is clear that these solutions are stable with
respect to 4-dimensional radial perturbations due to the fact that they 
are basically the static solution of a 1-dimensional analogous problem on 
the minimum of its potential.

{}Unlike the case of the previous section, in which the stability of the 
winding in the transverse direction is topological, the solutions obtained 
above are classically stable due to the dynamics in the transverse directions. 
They turn out to be stable if the radius $r$ of the string is
large enough. From the expansion (\ref{met2}) it is not difficult to see that 
the solution is stable   
under a small deviation in the $\tau$ direction, $\delta\tau(t)$, from 
the bottom of the conifold 
($\tau=0$). $\delta \tau$ contributes positive definite kinetic and potential 
energies to the action. Turning on a small perturbation $\delta \theta$ in 
the 
coordinate 
$\theta$, introduces a perturbation in the action (to second order in 
$\delta\theta$) proportional to
\begin{equation}
\left(\ell+{1\over 2}k^2(2n_\psi+n_\phi)^2\right)\dot{\delta\theta}^2
-{1\over 2k^2}\left(\ell-{1\over 2}k^2(2n_\psi+n_\phi)^2\right)\delta\theta^2~,
\end{equation}
with no mixing with the $\tau$ perturbations 
(whose lowest order is quadratic). We also notice that there is no contribution
to this order from perturbations in the other transverse directions. 
Therefore, the solution is stable as long as 
\begin{equation}
\ell>k^2(2n_\psi+n_\phi)^2/2~.
\end{equation}

{}Before we end this subsection, let us mention that a $\sigma$ dependence on 
the fluctuations considered does not change the result. Indeed, by considering 
\begin{equation}
  \delta\theta(\sigma,t)=\sum_n {\rm e}^{in\sigma}f_n(t)~,~~~~~~~f_n=f_{-n}~, 
\end{equation}
leads to an increase in the potential energy contribution of the $n$-th mode 
proportional to $n^2$, therefore improving the stability.

\subsection{D-string}

{}The action for a D-string is the sum of the Born-Infeld and Chern-Simons 
terms which, in the absence of a world-sheet gauge field, reads
\begin{equation}
S=-{\cal T}\int d\sigma dt~ 
{\rm e}^{-\Phi}\sqrt{-{\rm det}[\gamma+B_2]}+\mu\int C_2~,
\end{equation}
where the charge is related to the tension as $|\mu|={\cal T}$ and $C_2$ stands 
for the Ramond-Ramond two form. In the ansatz (\ref{ans}) for $\tau=0$ the equations
of motion coincide with those of the F-string since the contribution from 
the $F_3$ form vanishes in this case. However, the Chern-Simons term alters 
the stability condition under small $\theta$ fluctuations. It becomes:
\begin{equation}
\ell>{1\over2} k^2 \left[(2n_\psi+n_\phi)^2\pm (2n_\psi^2-n_\phi^2)\right]~,
\end{equation}
where $\pm=\mu/|\mu|$; while the stability under $\tau$ fluctuations is guaranteed for 
any choice of parameters as in the F-string case.

\subsection{Generalizations}

{}We expect the ansatz (\ref{ans}) to belong to wider class of stable classical
string solutions propagating at the bottom of the conifold geometry.  
In particular, we can also find solutions for the equations of motion for F and D strings
with arbitrary four dimensional shape if we wind the string along a maximal circle of the
$S^3$. In order to see that, it is sufficient to look at the general equations of motion 
in the conformal gauge for a string in the $\tau \rightarrow 0$ limit 
of our geometry. Also it is more convenient to use the familiar parametrization 
of the sphere, namely

\begin{equation}
ds^2=k^{-1}dx^\mu dx_\mu+ k\left(d\chi^2+\sin^2 \chi(d\theta^2 + \sin^2\theta d\phi^2)\right)~.
\end{equation}

If we now restrict ourselves to an ansatz of the form, $(t,x^i,0,\pi/2 ,\pi/2,\phi,0,0)$ 
then we see that the equations of motion (\ref{eom})
for the F and D string in this background are identical to the solutions found 
for the $M^4 \times S^1$ case
and therefore allow for any shape for the string in four dimensions. The stability of these
 solutions with respect to perturbations along the sphere 
follows from the arguments on previous sections.

\section{Concluding remarks}

We have found stable solutions for loops of F and D strings extended along the
uncompactified four dimensional space and winding a circle of the transverse 
3-sphere at the bottom of the warped deformed conifold. The reason for its 
stability from the four dimensional point of view is the modification of the 
string dynamics due to the massless degrees of freedom present on the string 
associated with the excitations along the transverse directions. This is 
reminiscent of what happens in the case of superconducting cosmic string case, 
where solutions of this type are known to exist, the so called vorton 
solutions \cite{Davis}. In fact, we can regard the strings propagating at the 
tip of this geometry as neutral superconducting strings, as long as the extra 
dimensional manifold there remains a 3-sphere. If this was the case on a 
realistic model of brane inflation, we should carefully study the cosmological 
implications of our findings. The existence of these stable loops that would not 
decay by gravitational radiation, would impose important constraints on the string 
network evolution \cite{Brandenberger}.

\section*{Acknowledgments}

{} We would like to thank Gia Dvali, Roberto Emparan, Gregory Gabadadze, Jaume Garriga, 
David Mateos, Ken Olum, Joseph Polchinski and Alexander Vilenkin for useful discussions.
J.J.B.- P. would like to thank the Departament of F\'{\i}sica Fonamental, at the Universitat 
de Barcelona for their hospitality during the early stages of this work. A.I. is supported
by funds provided by New York University and J.J.B.-P. is supported by the James Arthur Fellowship
at NYU.

\end{document}